\documentclass[twoside,american,english,prb,superscriptaddress,showpacs,amsmath]{revtex4}
\usepackage[T1]{fontenc}
\usepackage[latin9]{inputenc}
\usepackage{color}
\usepackage{amsmath}
\usepackage{amssymb}
\usepackage{graphicx}
\usepackage{esint}

\makeatletter
\@ifundefined{textcolor}{}
{%
 \definecolor{BLACK}{gray}{0}
 \definecolor{WHITE}{gray}{1}
 \definecolor{RED}{rgb}{1,0,0}
 \definecolor{GREEN}{rgb}{0,1,0}
 \definecolor{BLUE}{rgb}{0,0,1}
 \definecolor{CYAN}{cmyk}{1,0,0,0}
 \definecolor{MAGENTA}{cmyk}{0,1,0,0}
 \definecolor{YELLOW}{cmyk}{0,0,1,0}
 }

\tolerance7000\newcommand{\be}{\begin {equation}}\newcommand{\ee}{\end {equation}}\newcommand{\bear}{\begin{eqnarray}}\newcommand{\eear}{\end{eqnarray}}\topmargin0.1cm

\makeatother

\usepackage{babel}

\makeatother

\usepackage{babel}

\makeatother

\usepackage{babel}
\begin{document}

\title{Band Structure and Optical Absorption in Multilayer Armchair Graphene
Nanoribbons: A Pariser-Parr-Pople Model Study}

\author{Kondayya Gundra}

\altaffiliation{Permanent Address: Theoretical Physics Division, Bhabha Atomic Research Centre, Mumbai 400085, INDIA}

\email{naiduk@barc.gov.in, shukla@iitb.ac.in}

\author{Alok Shukla}

\affiliation{Department of Physics, Indian Institute of Technology, Bombay, Mumbai
400076 INDIA}
\begin{abstract}
Using the tight binding and Pariser-Parr-Pople (PPP) model Hamiltonians,
we study the electronic structure and optical response of multilayer
armchair graphene nanoribbons (AGNRs), both with and without a gate
bias. In particular, the influence of the number of layers ($n$),
and the strength of the electric field applied perpendicular to layers,
for different types of edge alignments, is explored on their electro-optical
properties. As a function of increasing $n$, the energy gap initially
decreases, eventually saturating for large $n$. The intensity of
the linear optical absorption in these systems also increases with
increasing $n$, and depends crucially on the polarization direction
of the incident light, and the type of the edge alignment. This provides
an efficient way of determining the nature of the edge alignment,
and $n$, in the experiments. In the presence of a gate bias, the
intensity of optical absorption behaves in a nontrivial way. The absorption
becomes more intense for the large fields in narrow ribbons exhibiting
a red shift of the band gap with the increasing field strength, while
in broad ribbons exhibiting a blue shift, the absorption becomes weaker.
However, for smaller electric fields, the absorption intensity exhibits
more complicated behavior with respect to the field strength. Thus,
the effect of the gate bias on optical absorption intensity in multilayer
AGNRs is in sharp contrast to the bilayer graphene, which exhibits
only enhancement of the absorption intensity with the increasing electric
field. 
\end{abstract}

\pacs{78.20.Bh, 78.67.Wj, 73.22.Pr , 78.40.Ri}

\maketitle

\section{Introduction\label{sec:intro} }

Since the isolation of graphene\cite{Novoselov-1}, rapid advances
have been made in the experimental\cite{Han} and theoretical\cite{graphene-rmp,peres-rmp,Ezawa}
investigations of this truly two dimensional (2D) material, and related
nanostructures, for use in the next generation opto-electronic devices.\cite{Sprinkel,LeiLiao}
While graphene is a zero gap semiconductor, energy gap opens up in
quasi one-dimensional (1D) graphene nanoribbons (GNRs) due to reduced
dimensions. The mechanism of gap opening in GNRs depends on the nature
of the edge termination. First principles calculations\cite{Son,QiminYan}
show that, in case of mono-layer zigzag GNRs (ZGNRs) energy gaps open
up due to edge magnetism. On the other hand, the gap opening in mono-layer
armchair GNRs (AGNRs), is believed to be both due to quantum confinement,
and the reduction of the bond lengths at the edges.\cite{Son} 

While numerous theoretical studies of the electronic structure and
related properties of the mono-layer GNRs exist,\cite{gnr-review}
relatively fewer calculations on bilayer and multilayer-GNRs have
been performed.\cite{mGNR1-Son1,mGNR2-Son2,mGNR3-sahu1,mGNR4-optics,mGNR5-edge-effects,mGNR6-edge-disorder,mGNR7-fast-slow-edges,mGNR8-pseudo-spin,mGNR9-Si(001),mGNR10-sahu2}
Recently, gated bilayer graphene has attracted a great deal of attention
in the experimental,\cite{Castro,Zhang-2,Lin,gated-bilayer-exp-prb,gated-bilayer-exp-stormer}
as well as theoretical communities.\cite{gated-bilayer-theory-mauri,gated-bilayer-theory-satpathy,gated-bilayer-theory-yang,gated-bilyayer-theory-sahu,gated-bilayer-theory-carbotte}
The energy gap in bilayer graphene, in the presence of a transverse
electric field, has been found to be tunable over a wide range of
values (up to 250 meV).\cite{Zhang-2,Castro} Furthermore, the electrical
noise levels in bilayer graphene channels is much less compared to
the noise levels in single layer graphene,\cite{Lin} thus making
bilayer graphene, a promising candidate for the fabrication of high-quality
electronic devices. Based upon this, one can argue that compared to
mono-layer GNRs, bilayer GNRs will possess superior properties from
the point of view of device physics. Moreover, it is interesting to
study multilayer graphene and GNRs in their own right, as they can
help us in understanding the evolution of the electronic structure
from graphene to bulk graphite. \cite{Peeters1}

Recently, we have developed an approach to study the electronic structure
of graphene nanostructures, based upon the $\pi$-electron Pariser-Parr-Pople
(PPP) model Hamiltonian,\cite{ppp} and used it to study the band
structure, edge magnetism, and optical absorption of mono -layer GNRs
of various types, at the Hartree-Fock level.\cite{gundra} The advantage
of the PPP model based methodology is that it incorporates the long-range
Coulomb interactions among the $\pi$-electrons in a natural way.
In this work, we apply our approach\cite{gundra} to perform a detailed
investigation of the electronic structure of multilayer AGNRs, with
and without an external electric field, and for various edge alignments.
For the gated ribbons with the intrinsic band gaps above a critical
value (say, $\epsilon_{c}$), the gap decreases with the increasing
field strength, while for those with gaps lower than $\epsilon_{c}$,
it increases with the external field, a result in good agreement with
the recent \emph{ab initio} results of Sahu \emph{et al.}\cite{mGNR10-sahu2}
In addition to the studies of the band structure, we also present
calculations of the optical absorption spectra of various multilayer
AGNRs for different edge alignments, for various polarization directions
of the incident light, and with, and without, a gate bias.\textcolor{magenta}{{}
}\textcolor{black}{To the best of our knowledge, no prior calculations
of optical absorption in multilayer-GNRs exist, which take electron-electron
interactions into account. Our calculations reveal that for the gated
bilayer ribbons, for large bias fields, the optical absorption intensity
increases for the ribbons with the intrinsic band gaps higher than
$\epsilon_{c}$, while for the ribbons with the gaps smaller than
$\epsilon_{c}$, the absorption intensity decreases. However, for
smaller bias fields, the absorption intensity exhibits a more complicated
behavior with respect to the field strength, a behavior in sharp contrast
with the gated bilayer graphene which exhibits only increase in the
absorption intensity with the increasing bias field.}\cite{Zhang-2,Castro,gated-bilayer-theory-yang}
Furthermore, we find that in addition to the polarization direction,
the absorption intensity is found to be crucially dependent on the
number of layers in the multilayer AGNRs, and their edge alignment,
thereby allowing for optical determination of their structure.

The remainder of this paper is organized as follows. In section \ref{sec:PPP-Theory},
we briefly describe our PPP model based theoretical methodology. Next,
in Sec. \ref{sec:band} we present the results of our calculations
on the band structure of various multilayer AGNRs, and discuss the
variation of the band gap with the increasing number of layers, their
widths, and as a function of the gate bias. In section \ref{sec:optical}
we present the optical absorption spectra of these systems, and discuss
their variation with $n$, external electric field, and the type of
edge alignment. Finally, in Sec \ref{sec:summary} we summarize our
results, and present the conclusions.

\section{Theoretical Methodology \label{sec:PPP-Theory}}

In our earlier works we have used the PPP model to extensively to
study the electronic structure and optical properties of \emph{finite}
$\pi$-electron systems such as conjugated molecules and oligomers
at various levels of theory,\cite{shukla} while in our recent work
we extended it to study GNRs in the infinite length limit, with 1D
periodic boundary conditions imposed.\cite{gundra} The PPP model
Hamiltonian,\cite{ppp} with one $\pi$-electron per carbon atom,
is given by 
\begin{eqnarray}
H=-\sum_{i,j,\sigma}t_{ij}(c_{i\sigma}^{\dagger}c_{j\sigma}+c_{j\sigma}^{\dagger}c_{i\sigma})+\nonumber \\
U\sum_{i}n_{i\uparrow}n_{i\downarrow}+\sum_{i<j}V_{ij}(n_{i}-1)(n_{j}-1)\label{eq:ham-ppp}
\end{eqnarray}
 where $c_{i\sigma}^{\dagger}$ creates an electron of spin $\sigma$
on the $p_{z}$ orbital of carbon atom $i$, $n_{i\sigma}=c_{i\sigma}^{\dagger}c_{i\sigma}$
is the number of electrons with the spin $\sigma$, and $n_{i}=\sum_{\sigma}n_{i\sigma}$
is the total number of electrons on atom $i$. The parameters $U$
and $V_{ij}$ are the on-site and long-range Coulomb interactions,
respectively, while $t_{ij}$ is the one-electron hopping matrix element.
On setting $U=V_{ij}=0$, the Hamiltonian reduces to tight binding
(TB) model, while on setting just $V_{ij}=0$, it reduces to the Hubbard
model. The parametrization of Coulomb interactions is Ohno like,\cite{Ohno}
\begin{equation}
V_{i,j}=\frac{U}{\kappa_{i,j}(1+0.6117R_{i,j}^{2})^{1/2}}\;\mbox{,}\label{eq-ohno}
\end{equation}
where, $\kappa_{i,j}$ depicts the dielectric constant of the system
which can simulate the effects of screening, and $R_{i,j}$ is the
distance in \AA{}~ between the $i$-th and the $j$-th carbon atoms.
In our earlier work on GNRs, \cite{gundra} we used the \emph{ab-initio}
GW band structure of mono layer AGNR-12 (AGNR-$N_{A}$, denotes an
AGNR with $N_{A}$ dimer lines across the width) reported by Son \emph{et
al.}\cite{Son} to obtain a set of {}``modified screened Coulomb
parameters,'' with $U=6.0$ eV and $\kappa_{i,j}=2.0$ ($i\neq j)$
and $\kappa_{i,i}=1$. Note that our modified screened parameters
are slightly different from the screened parameters reported initially
by Chandross and Mazumdar,\cite{chandross} with $U=8.0$ eV and $\kappa_{i,j}=2.0$
($i\neq j)$ and $\kappa_{i,i}=1$, aimed at describing the optical
properties of phenyl-based polymers within the PPP model. With our
modified screened parameters, we obtained excellent agreement between
the\emph{ ab initio} GW band gaps,\cite{Son} and our PPP Hartree-Fock
band gaps for mono layer AGNRs of various widths,\cite{gundra} therefore,
we have used these parameters in the present study of multilayer AGNRs
as well. 

\begin{figure}

\includegraphics[width=8.2cm]{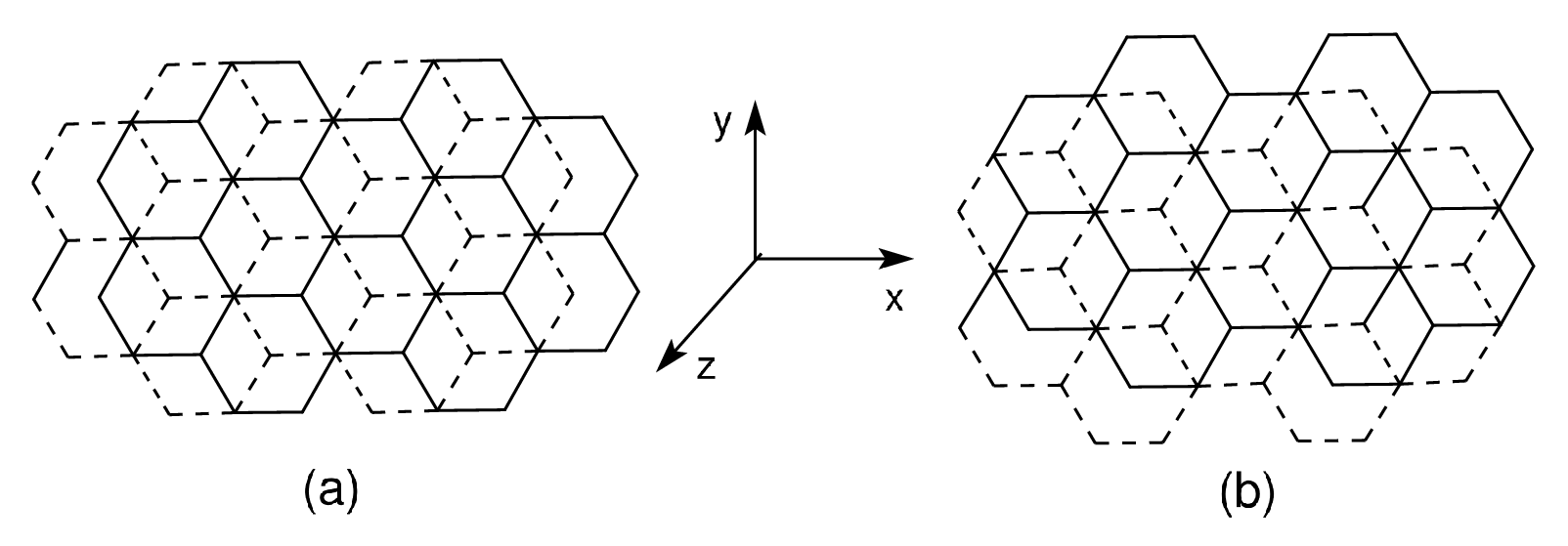}
\caption{The structures of bilayer AGNR (a) $\alpha$-alignment and (b) $\beta$-alignment }

\label{Fig:alpha-beta}
\end{figure}

In these calculations, we consider Bernal packed multilayer AGNRs
with ABAB... repeat pattern, and two possible edge alignments, called
$\alpha$ and $\beta$ alignments,\cite{mGNR3-sahu1} shown in Fig.
\ref{Fig:alpha-beta}. The intra layer nearest-neighbor (NN) distance
was taken to be 1.42 \AA, and, for the inter layer separation the
value 3.35 \AA, identical to that in graphite, was used. As far as
the hoppings are concerned, the intra layer NN hopping was chosen
to be $t=2.7$ eV, along with the next-nearest-neighbor (NNN) hopping
$t^{'}=0.27$ eV. For the inter-layer hopping also, we considered
not only NN hopping $t_{\bot}=0.4$ eV, but also the NNN hopping integral
$t_{\bot}^{'}=0.3$ eV.\cite{graphene-rmp} We further explore the
parameter dependence of the calculated optical absorption spectra
in section \ref{sub:parameter-values}. Because multilayer AGNRs are
closed-shell insulating systems, like their mono layer counterparts,
we used the restricted HF (RHF) method to obtain the band structure
in our calculations, as described in our previous work.\cite{gundra} 

\section{energy gaps and band structure }

\label{sec:band}

\subsection{Energy gaps }

\label{sub:gaps}

\begin{figure}
\vspace*{3mm}
\includegraphics[width=6cm]{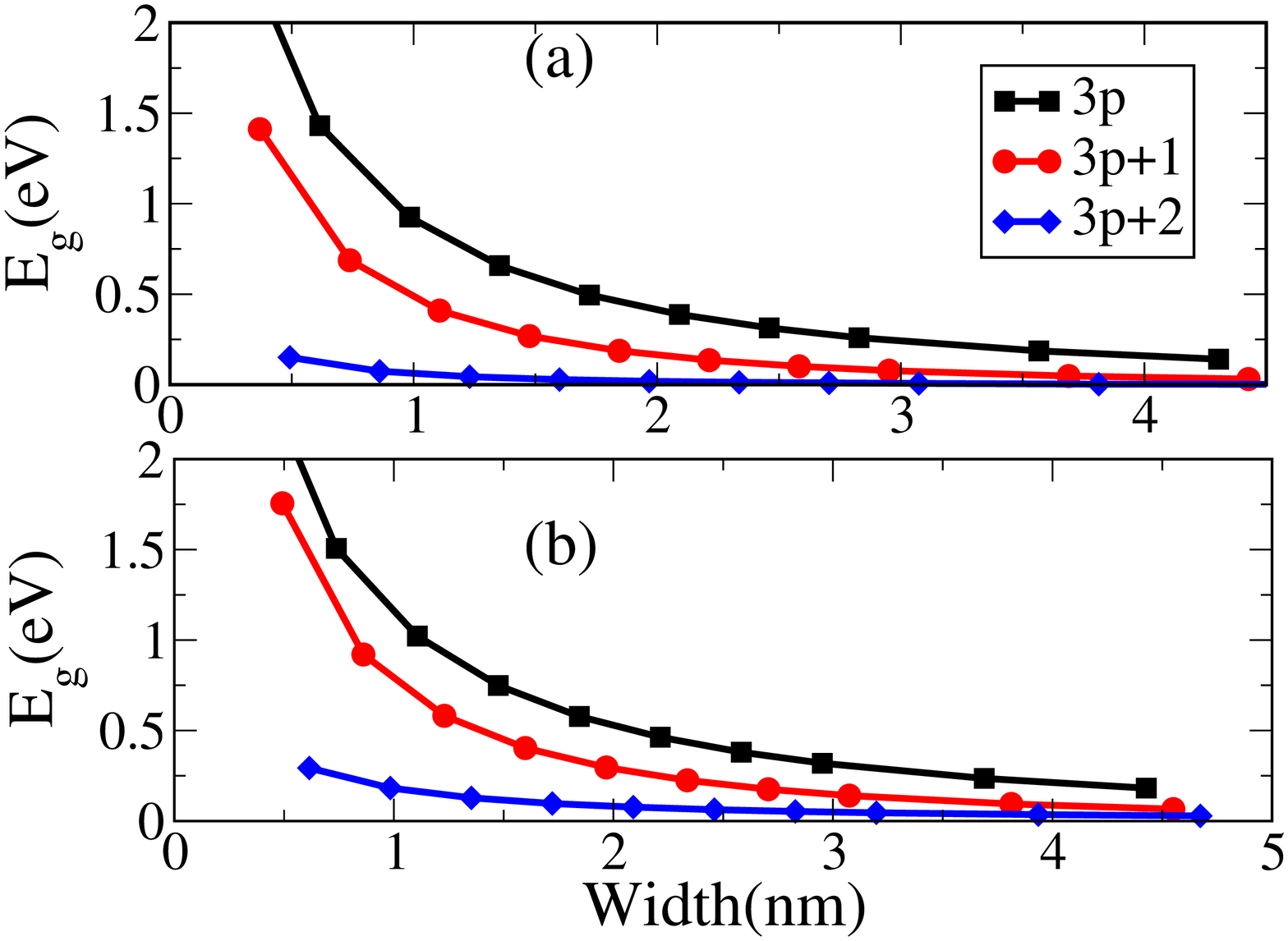}

\caption{(Color online) Variation of the energy gap with the increasing width
for bilayer AGNRs, obtained by the tight-binding method for (a) $\alpha$-alignment
(b) $\beta$-alignment.}

\label{Flo:eg-width-tb-2agnr}
\end{figure}

\begin{figure}
\includegraphics[width=6cm]{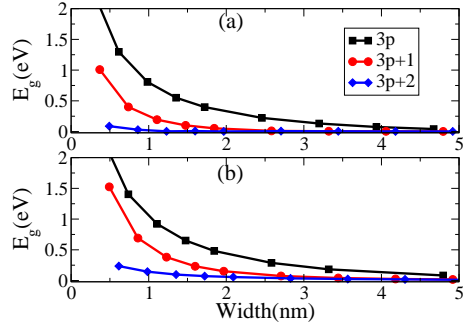}

\caption{(Color online) Variation of the energy gap with the increasing width
for the AGNRs with eight layers, obtained by the tight-binding method
for (a) $\alpha$-alignment (b) $\beta$-alignment.}

\label{Flo:eg-width-tb-8agnr}
\end{figure}

The TB calculations on mono-layer AGNRs\cite{Nakada} predicted that
AGNR$-N_{A}$ with $N_{A}=3p$ and $N_{A}=3p+1$ ($p$ is a positive
integer) exhibit energy gaps ($E_{g}^{N_{A}}$), which are inversely
proportional to $N_{A}$, whereas AGNRs with $N_{A}=3p+2$ are gapless.
Hence, based upon TB theory, they can be classified into three groups
with energy gaps varying as $E_{g}^{3p}\geq E_{g}^{3p+1}\geq E_{g}^{3p+2}(=0)$.
However, \emph{ab initio} density-functional theory (DFT) calculations\cite{Son}
on these ribbons predicted a different relationship $E_{g}^{3p+1}\geq E_{g}^{3p}\geq E_{g}^{3p+2}(\neq0)$,
with the important result that even for $N_{A}=3p+2$, AGNRs exhibit
nonzero energy gaps, due to the fact that the bond lengths involving
the edge atoms are shorter than those in the interior. When the decrease
in the bond length is incorporated in the TB approach by increasing
the corresponding hopping, one also obtains finite gaps for $N_{A}=3p+2$
ribbons, although the relation $E_{g}^{3p}\geq E_{g}^{3p+1}\geq E_{g}^{3p+2}$
still holds. Based upon first principles DFT calculations Sahu \emph{et
al.,} \cite{mGNR3-sahu1,mGNR10-sahu2} discovered that for bilayer
AGNRs, and other multilayer AGNRs $E_{g}^{3p+1}\geq E_{g}^{3p}\geq E_{g}^{3p+2}(\neq0)$
is valid, in full agreement with the DFT results of Son \emph{et al.}\cite{Son}
obtained for the mono-layer AGNRs. In what follows we first explore
the same relationship using the TB model, and in Figs. \ref{Flo:eg-width-tb-2agnr}
and \ref{Flo:eg-width-tb-8agnr} we present the variation of the energy
gaps of different families of bilayer and eight layer AGNRs, respectively,
with respect to their widths, both for the $\alpha$ and the $\beta$
alignments. 
The values of \textcolor{black}{of the hopping integrals used in these
calculations were as specified in section}\textcolor{red}{{} }\textcolor{black}{\ref{sec:PPP-Theory}},
except that the hoppings at the edges were increased by 12\%,\cite{Son}
leading to finite band gaps for ribbons with $N_{A}=3p+2$ for the
narrow multilayer ribbons. As is obvious from the figure, our results
are consistent with the aforesaid relation $E_{g}^{3p}\geq E_{g}^{3p+1}\geq E_{g}^{3p+2}$
valid for mono-layer AGNRs at the TB level. Furthermore, for a given
width of the bilayer and \textcolor{black}{eight layer} AGNRs, band
gaps with the $\beta$ alignment are always more than that for the
$\alpha$ alignment. 

\begin{figure}
\vspace*{3mm}
\includegraphics[width=6cm]{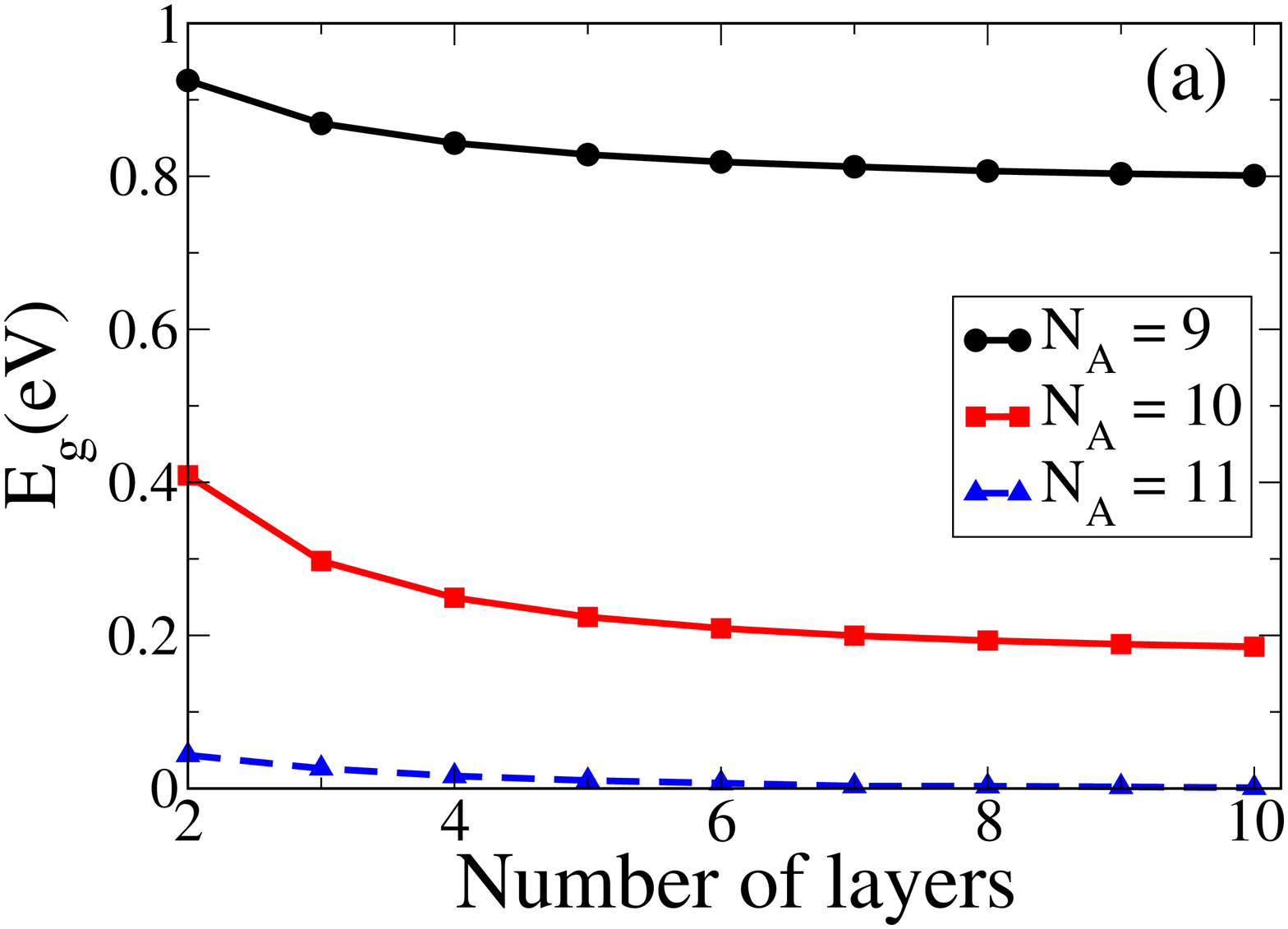}
\vspace*{3mm}
\includegraphics[width=6cm]{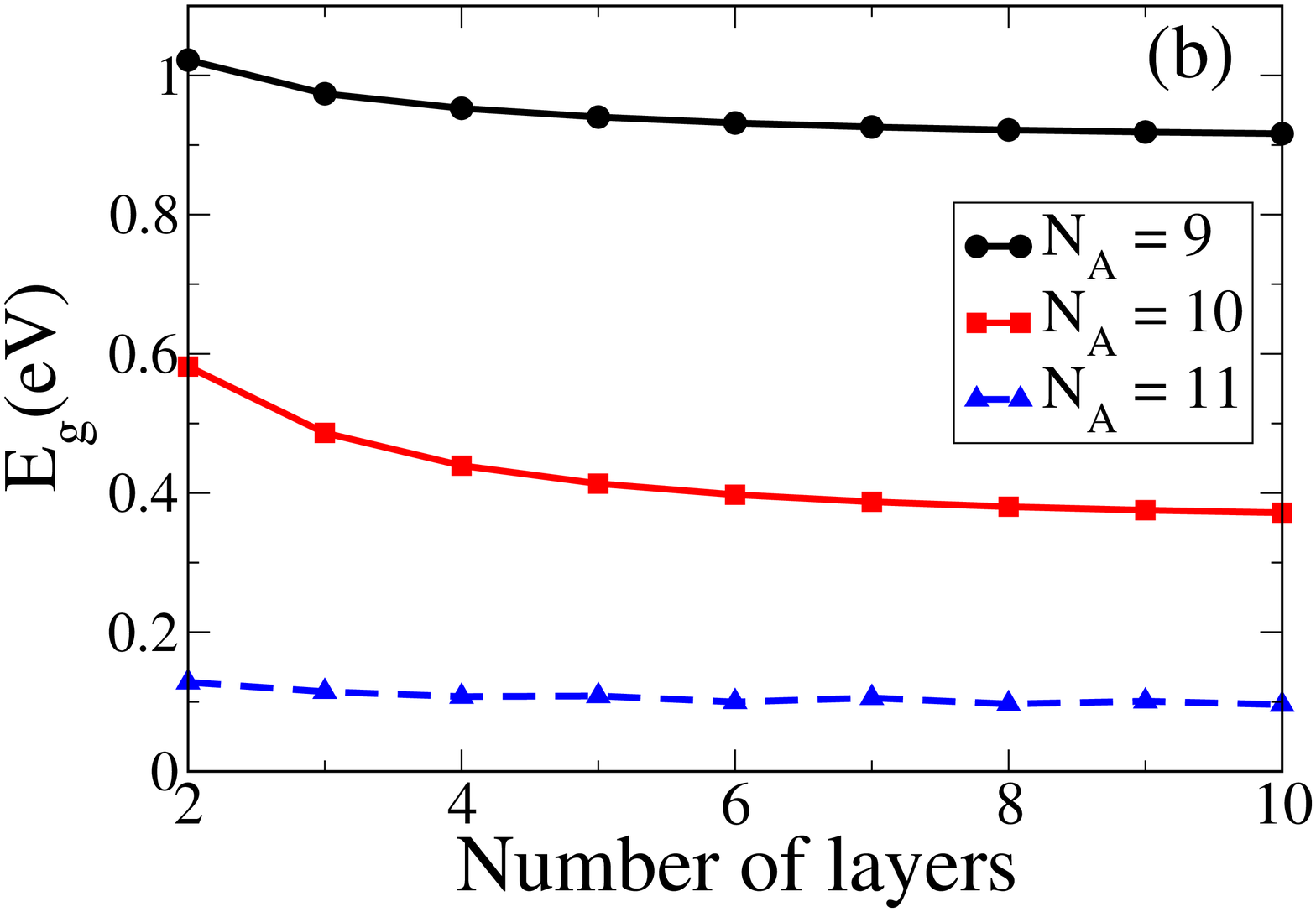}

\caption{(Color online) Variation of energy gap with number of layers obtained
by the tight-binding method for multilayer AGNRs in (a) $\alpha$-alignment
(b) $\beta$-alignment.}

\label{Fig:gap-nl-3p12}
\end{figure}

Next, for each of the three families of AGNRs, we examine the variation
of the gap as a function of the increasing number of layers $(n)$
at the TB level, and results of our calculations are presented in
Fig. \ref{Fig:gap-nl-3p12}a for the $\alpha$ alignment, and in Fig.
\ref{Fig:gap-nl-3p12}b for the $\beta$ alignment. Although, these
calculations were performed for $p=3$, a similar behavior was also
observed for other values of $p$. We find that: (a) again the band
gaps for the ribbons in $\beta$-alignment are larger than those for
the $\alpha$-alignment, and (b) band gaps for both types of alignments
saturate fairly rapidly with respect to $n$, suggesting that a small
number of layers are needed to achieve the bulk values. 

\begin{figure}
\includegraphics[width=6cm]{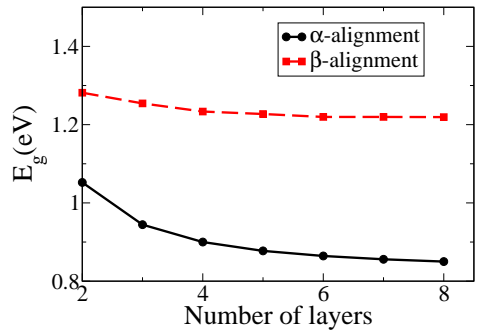}

\caption{(Color online) Variation of energy gap with number of layers obtained
using the PPP model, within the RHF approximation, for the multilayer
AGNRs with $N_{A}=8$, in $\alpha$ and $\beta$ alignments.}

\label{Flo:eg-nl-ppp}
\end{figure}

In particular, for the $\alpha$-aligned ribbons with $N_{A}=11$
\textcolor{black}{(}\foreignlanguage{american}{\textcolor{black}{$3p+2$}}\textcolor{black}{{}
class)} , the band gap decreases rapidly with $n$ and becomes negligibly
small for large values of $n$ (Fig.\ref{Fig:gap-nl-3p12}(a), blue/dashed
line) whereas, for the $\beta$ alignment the band gap is much less
sensitive to $n$ and attains a constant value, for smaller values
of $n$ (Fig.\ref{Fig:gap-nl-3p12}(b), blue/dashed line). This tendency
holds even when the e-e interactions are considered, as is evident
in the Fig. \ref{Flo:eg-nl-ppp} which presents the variation of band
gap with number of layers for multilayer AGNRs with $N_{A}=8$ in
$\alpha$-alignment (black line) and in $\beta$-alignment (\textcolor{black}{red
line}), obtained by PPP-RHF approach. From the results it is obvious
that the inclusion of the e-e interactions leads to the widening up
of the band gaps, compared to the TB approach. 

\subsection{Band structure}

\label{sub:bands}

\begin{figure}
\includegraphics[width=5cm]{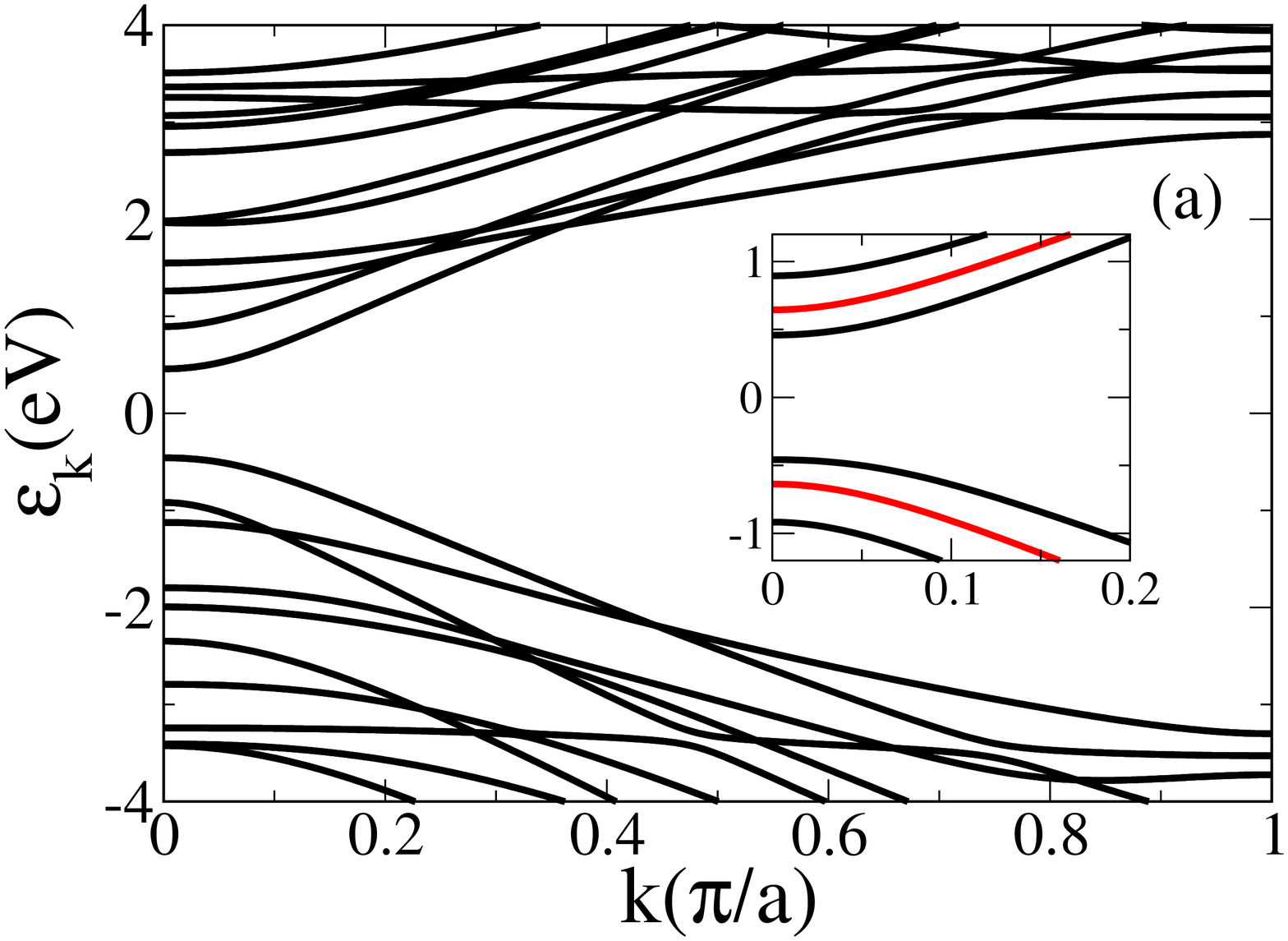} \includegraphics[width=5cm]{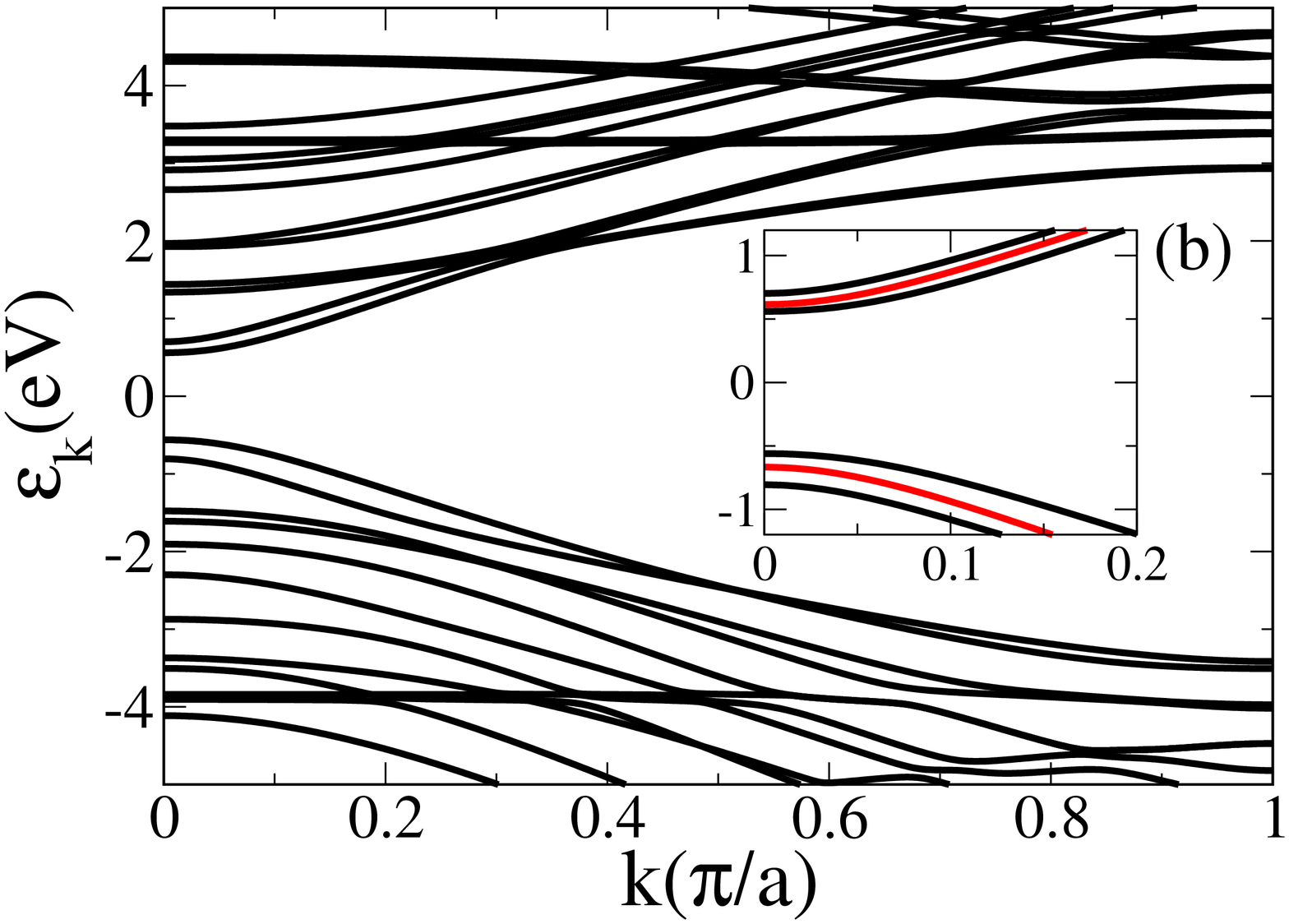}

\caption{(Color online) Band structure of (a) 2-AGNR-11-$\alpha$ , (b) 2-AGNR-11-$\beta$,
obtained using the PPP-RHF approach. The inset contains the magnified
band structure near the Fermi energy ($E_{F}=0$) of: (a) 2-AGNR-11
(black lines) and (b) mono-layer AGNR-11 (red lines).}

\label{Fig:band-ba11-alpha-beta}
\end{figure}

\begin{figure}
\includegraphics[width=8cm]{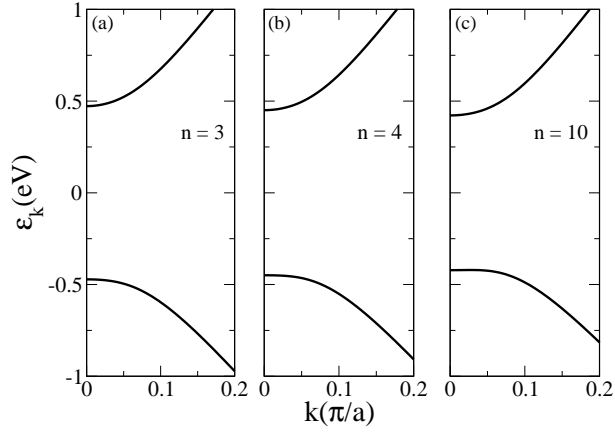} 

\caption{Band structure of (a) 3-AGNR-8-$\alpha$ , (b) 4-AGNR-8-$\alpha$,
(c) 10-AGNR-8-$\alpha$ near $E_{F}$ obtained using the PPP-RHF approach.}

\label{Fig:band-3-10-agnr-alpha}
\end{figure}

For the sake of brevity, in what follows, we denote a given multilayer
AGNR as $n$-AGNR-$N_{A}$-$\alpha$ ($n$-AGNR-$N_{A}$-$\beta$),
implying an AGNR with $n$ layers, each of width $N_{A}$, arranged
in $\alpha$ ($\beta$) alignment. In our previous work we have emphasized
the role of long-range e-e interactions in widening the band gaps
of the mono-layer AGNRs.\cite{gundra} For example, we demonstrated
that for mono-layer AGNR-11, the Hubbard model based calculations
predict a negligible gap, while the PPP-RHF method predicts a gapped
system in agreement with the GW calculations.\cite{Son} We expect
the long-range e-e interactions, as incorporated in the PPP model,
to play a similar role for multilayer AGNRs as well. \textcolor{black}{According
to the }\textcolor{black}{\emph{ab initio}}\textcolor{black}{{} DFT
calculations of Sahu }\textcolor{black}{\emph{et al.}}\textcolor{black}{\cite{mGNR3-sahu1,mGNR10-sahu2},
all $n$-AGNR-$N_{A}$-$\alpha$, of the family $N_{A}=3p+2$, exhibit
much smaller gaps compared to the other families, suggesting a metal-like
behavior.} In Figs. \ref{Fig:band-ba11-alpha-beta}a and \ref{Fig:band-3-10-agnr-alpha}
we present the PPP-RHF band structures of a few members of this family,
2-AGNR-11-$\alpha$, 3-AGNR-8-$\alpha$, 4-AGNR-8-$\alpha$, and 10-AGNR-8-$\alpha$,
and from the figures it is obvious that all these ribbons have substantial
band gaps. \textcolor{black}{Therefore, we believe that the }\textcolor{black}{\emph{ab
initio}}\textcolor{black}{{} results of Sahu }\textcolor{black}{\emph{et
al.}}\textcolor{black}{\cite{mGNR3-sahu1,mGNR10-sahu2} reporting
these systems as almost metallic,} are due to the well-known tendency
of the DFT to underestimate the gaps, and that a better estimate of
the gap can only be made by some electron-correlated approach such
as the GW approximation.\cite{Son} 

We investigate the effect of edge alignment on the band structure
of bilayer AGNRs by presenting the band structure of 2-AGNR-11-$\alpha$
(Fig.\ref{Fig:band-ba11-alpha-beta} (a)), and 2-AGNR-11-$\beta$
(Fig.\ref{Fig:band-ba11-alpha-beta}(b)). The band structure near
the Fermi energy is magnified in the insets of those figures, which
also contain the band structure of mono-layer AGNR-11. In the presence
of the second layer, each mono-layer band is split into two bands:
one with the lower energy and the other with the higher energy, thus
reducing the gap. From insets of Figs. \ref{Fig:band-ba11-alpha-beta}a
and \ref{Fig:band-ba11-alpha-beta}b, it is clear that the \textcolor{black}{valence
and the conduction} bands formed due to the aforesaid band splitting
for 2-AGNR-11-$\alpha$ near $E_{F}$ are separated by larger energy
compared to the case of 2-AGNR-11-$\beta$. Therefore, it suggests
that the perturbation introduced by the second layer is larger in
the case of the $\alpha$-aligned AGNRs, as compared to the $\beta$-aligned
AGNRs, for which we offer the following geometrical explanation. Intuitively
speaking, the layers will interact with each other the most if they
were stacked in the AA arrangement (with all the carbon atoms on top
of each other), leading to a more prominent band splitting, and thus
the smallest gap possible from various stacking arrangements. For
the AB stacking, however, in the $\alpha$ case, the difference with
respect to the AA stacking is much less because the layers are only
displaced in the $x$ direction, while in the $\beta$ case, layers
are displaced both in the $x$ as well as the $y$ directions, resulting
in more disalignment compared to the $\alpha$ case (\emph{cf}. Fig.
\ref{Fig:alpha-beta}). Therefore, in $\alpha$ alignment the layers
will interact with each other more than those in the $\beta$ case,
resulting in a band structure more distinct compared to a mono-layer,
and, thus, a smaller gap, compared to the $\beta$ case. We verified
this hypothesis by actually performing the band structure calculations
for 2-AGNR-11 in different stacking patterns and alignments, and the
band gaps (in eV) obtained by using the TB/PPP-RHF method were 0.02/0.43
(AA), 0.04/0.92 (AB-$\alpha$) and 0.13/1.12 (AB-$\beta$). 

Next, we illustrate the variation in the band structure with the number
of layers ($n$), by presenting the band structure near $E_{F}$ of
3-AGNR-8-$\alpha$ (Fig.\ref{Fig:band-3-10-agnr-alpha}a), 4-AGNR-8-$\alpha$
(Fig.\ref{Fig:band-3-10-agnr-alpha}b), and 10-AGNR-8-$\alpha$ (Fig.\ref{Fig:band-3-10-agnr-alpha}c).
The band structure changes significantly with increasing $n$, \textcolor{black}{leading
not just to more bands}, but also causing bands near $E_{F}$ to become
flatter. Furthermore, this flatness extends more and more into BZ
as $n$ increases, which shows up as a tremendous increase in the
joint density of states not just due to a denser band structure, but
also due to van Hove singularity like increase caused by parallel
bands. This has obvious implications for the optical absorption spectra
of such ribbons, which will be discussed further in Sec. \ref{sub:opt-layers}. 

\subsection{Effect of gate bias on the electronic structure}

\begin{figure}
\includegraphics[width=6cm]{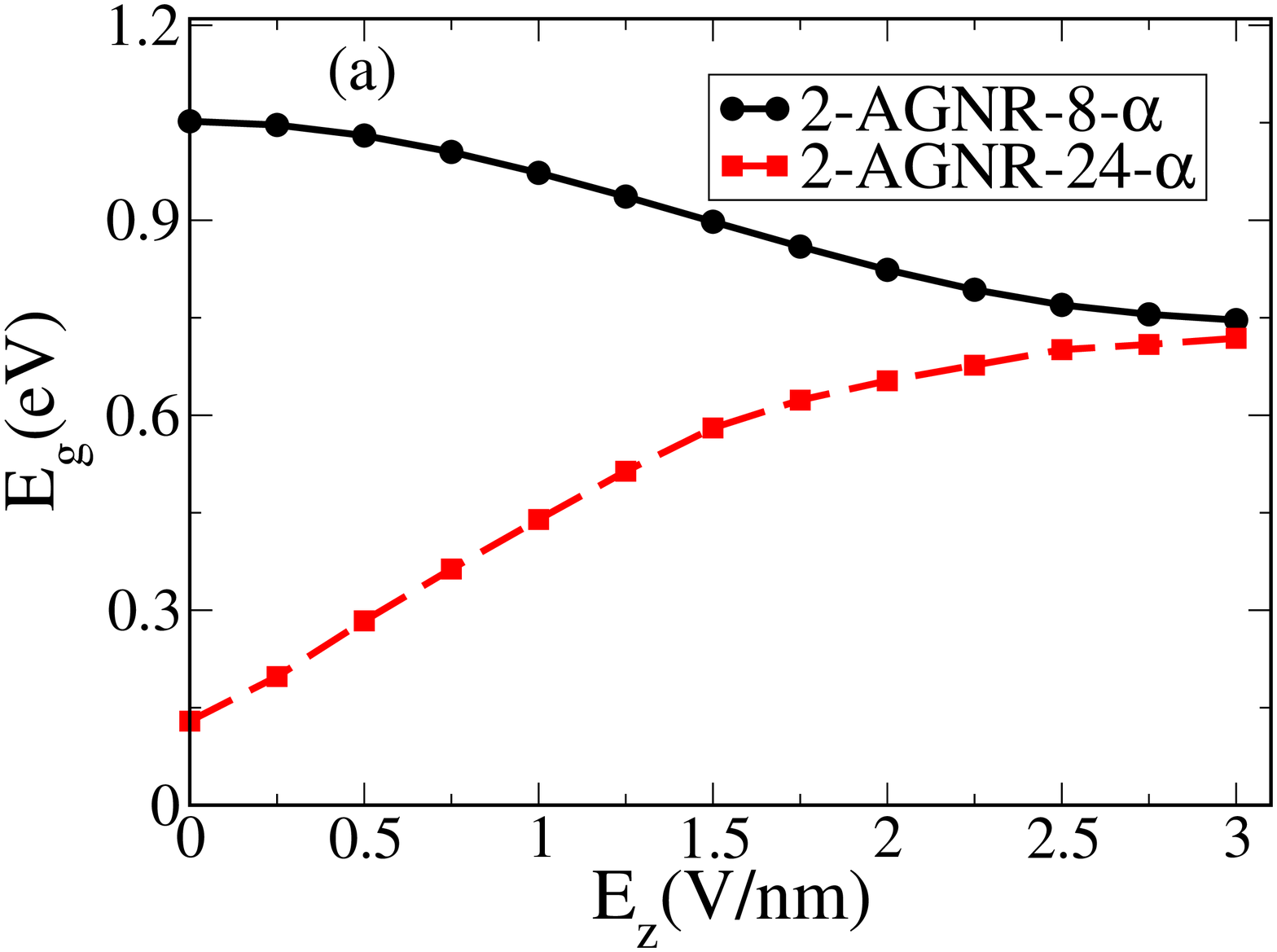}

\includegraphics[width=6cm]{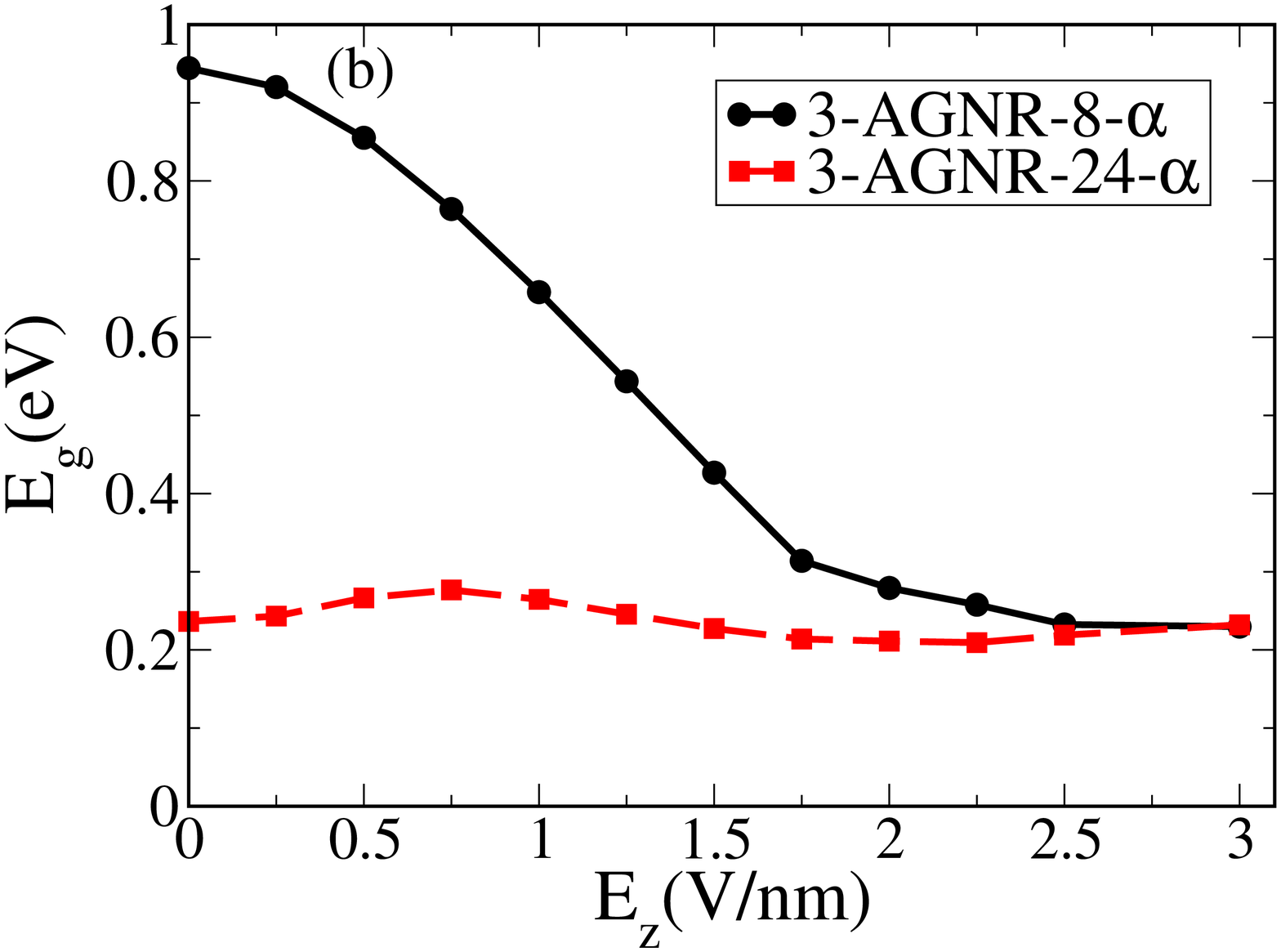}

\caption{(Color online) Variation of energy gap with $E_{z}$ for (a) 2-AGNR-8-$\alpha$
and 2-AGNR-24-$\alpha$ (b) 3-AGNR-8-$\alpha$ and 3-AGNR-24-$\alpha$,
obtained using the PPP-RHF approach.}

\label{Fig:Eg-Ez}
\end{figure}

The variation of band gaps of bilayer AGNRs, when electric ($E_{z}$)
is applied along $z$-direction is quite interesting. \emph{Ab initio}
DFT study on these systems by Sahu \emph{et al.,\cite{mGNR3-sahu1}}
revealed that the band gaps increase with the increasing $E_{z}$
for ribbons with band gaps below a critical value, and decrease for
ribbons with band gaps above this critical value. Using nearest neighbor
TB approach, coupled with perturbation theory, they showed that for
the $\alpha$-aligned ribbons, the critical band gap $\epsilon_{c}=(\sqrt{5}-1)t_{\perp}$,
so that for $t_{\bot}=0.4$, i $\epsilon_{c}\approx0.25$ eV. In Fig.
\ref{Fig:Eg-Ez}a we present the variation of band for 2-AGNR-8-$\alpha$
and 2-AGNR-24-$\alpha$ with $E_{z}$, obtained using the PPP-RHF
calculations. In the absence of $E_{z}$, the band gap of 2-AGNR-8-$\alpha$
is larger than the $\epsilon_{c}$, hence the band gap decreases with
$E_{z}$ (Fig. \ref{Fig:Eg-Ez}a, black solid line), whereas, the
band gap of 2-AGNR-24-$\alpha$ in the absence of $E_{z}$ is less
than $\epsilon_{c}$, \textcolor{black}{hence the band gap increases
with $E_{z}$ (Fig. \ref{Fig:Eg-Ez}a, red lines).} In Fig. \ref{Fig:Eg-Ez}b
we present the variation of band gap for 3-AGNR-8-$\alpha$ and 3-AGNR-24-$\alpha$
with $E_{z}$, an\textcolor{black}{d we observe a trend similar to
the case of the bilayer GNRs} for 3-AGNR-8-$\alpha$, but for 3-AGNR-24-$\alpha$
the gap exhibits small oscillations around a uniform value. The trends
on the variation of band gaps with $E_{z}$, obtained from our PPP-RHF
calculations are fully consistent with the \emph{ab initio }DFT results.\cite{mGNR3-sahu1,mGNR10-sahu2} 

\begin{figure}
\includegraphics[width=8cm]{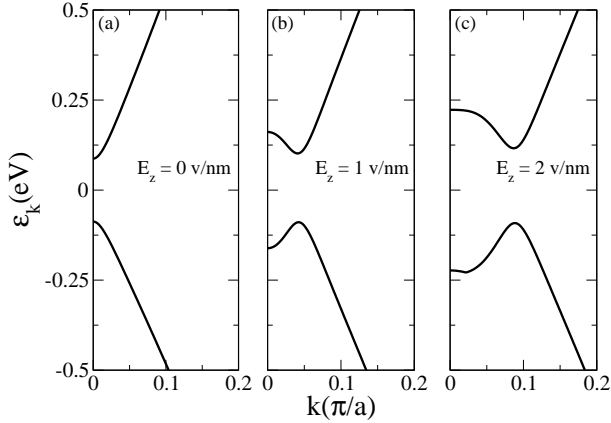}

\caption{PPP-RHF band structure of 3-AGNR-24-$\alpha$ near $E_{F}$, for several
values of external electric field $E_{z}$}

\label{Flo:band-3agnr-ez}
\end{figure}

Finally, we examine the variations in the band structures of multilayer
AGNRs caused by the gate bias. In the absence of an external electric
field, the fundamental gap of all the ribbons is located at $k=0$.
However, when the electric filed is applied, location of the fundamental
gap shifts in the BZ, from $k=0$ to finite values of $k$, \textcolor{black}{similar
to what was observed in the gated multi-layer graphene as well.\cite{Peeters2}}
The rightward shift of fundamental gap continues with the increasing
$E_{z}$ which is evident from the Fig. \ref{Flo:band-3agnr-ez} ,
where we have presented the band structure of of 3-AGNR-24-$\alpha$
for increasing values of $E_{z}$. 

\section{optical absorption spectra \label{sec:optical}}

\subsection{Basic formalism}

The optical absorption spectrum for incident radiation polarized in
$x$ or $y$ or $z$ direction is computed in the form of the corresponding
components of the imaginary part of the dielectric constant tensor,
\emph{i.e}., $\epsilon_{ii}(\omega)$ using the standard formula
\begin{equation}
\epsilon_{ii}(\omega)=C\sum_{v,c}\int_{-\pi/a}^{\pi/a}\frac{|\langle c(k)|p_{i}|v(k)\rangle|^{2}}{\{(E_{cv}(k)-\hbar\omega)^{2}+\gamma^{2}\}E_{cv}^{2}(k)}dk,\label{eq:eps2}
\end{equation}
where $a$ is the 1D lattice constant, $|c(k)\rangle$ ($|v(k)\rangle$)
denotes conduction (valence) band state, $p_{i}$ denotes the momentum
operator in the $i$-th Cartesian direction, $\omega$ represents
the angular frequency of the incident radiation, $E_{cv}(k)=\epsilon_{c}(k)-\epsilon_{v}(k)$,
with $\epsilon_{c}(k)\:(\epsilon_{v}(k))$ being the conduction (valence)
band eigenvalues of the Fock matrix, $\gamma$ is the line width,
while $C$ includes rest of the constants. As in our previous work,\cite{gundra}
the momentum matrix elements $\langle c(k)|p_{i}|v(k)\rangle$ were
computed numerically using the formula proposed by Pedersen \emph{et
al}.\cite{opt-mat-el-2}

\subsection{\textcolor{black}{Sensitivity of the optical absorption to Hamiltonian
parameters}}

\textcolor{black}{\label{sub:parameter-values}}

\begin{figure}
\includegraphics[width=6cm]{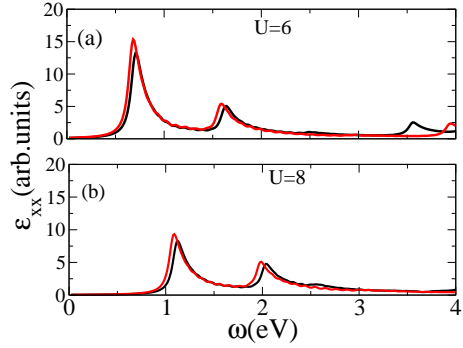}\label{Fig:sig-x-ba11-para}

\caption{(Color online) Optical absorption spectra of 2-AGNR-11-$\alpha$ calculated
using the PPP-RHF approach for the light polarized along the $x$
axis with Coulomb parameters (a) $U=6.0$ eV and $\kappa_{i,j}=2.0$
($i\neq j)$ and $\kappa_{i,i}=1$ and (b) $U=8.0$ eV and $\kappa_{i,j}=2.0$
($i\neq j)$ and $\kappa_{i,i}=1$. In both the cases the black lines
correspond to hopping parameters (in eV) $t=2.7$, $t^{'}=0.27$,
$t_{\bot}=0.4$ eV, and $t_{\bot}^{'}=0.3$, the red line corresponds
to $t=3.16$,$t^{'}=0.316$, $t_{\bot}=0.39$ eV, and $t_{\bot}^{'}=0.315$. }
\end{figure}

\begin{figure}
\vspace*{3mm}
\includegraphics[width=6cm]{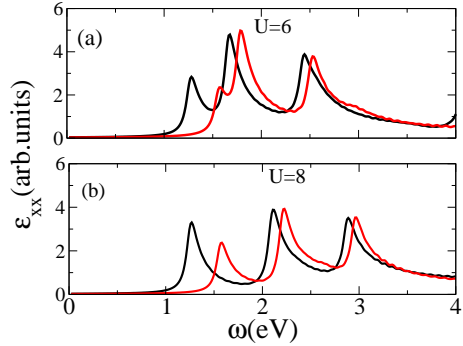}\label{Fig:sig-x-ba12-para}

\caption{(Color online) Optical absorption spectra of 2-AGNR-12-$\alpha$ computed
using the PPP-RHF approach for the light polarized along the $x$
axis, with various sets of PPP parameters as explained in the caption
of Fig. \ref{Fig:sig-x-ba11-para}.}
\end{figure}

\textcolor{black}{Before we present and discuss our results on the
optical absorption in multilayer ANGRs, we examine their sensitivity
to the hopping and the Coulomb parameters used in the PPP model Hamiltonian.
We used two sets of Coulomb parameters: $U=6.0$ eV and $\kappa_{i,j}=2.0$
($i\neq j)$ and $\kappa_{i,i}=1$ (our modified screened parameters\cite{gundra})
and $U=8.0$ eV and $\kappa_{i,j}=2.0$ ($i\neq j)$ and $\kappa_{i,i}=1$
(original screened parameters of Chandross and Mazumdar\cite{chandross}).
Similarly, for the hoppings we used two sets: commonly used values
of $t=2.7$ eV ,$t^{'}=0.27$ eV, $t_{\bot}=0.4$ eV, $t_{\bot}^{'}=0.3$
eV, and a set with larger values of hoppings, }\textcolor{black}{\emph{viz}}\textcolor{black}{.,
$t=3.16$,$t^{'}=0.316$, $t_{\bot}=0.39$ eV, and $t_{\bot}^{'}=0.315$.\cite{tb-parameters}
In Figs. \ref{Fig:sig-x-ba11-para} and \ref{Fig:sig-x-ba12-para}
we present the results of these calculations for the ribbons 2-AGNR-11-$\alpha$
and 2-AGNR-12-$\alpha$, respectively. From the figures the following
trends emerge: (a) For 2-AGNR-11-$\alpha$, the change of hoppings
(red line vs black line) does not change the results significantly,
but the increase in the value of $U$ ($U=6$ vs. $U=8$) leads to
a prominent blue shift in the spectrum, (b) For 2-AGNR-12-$\alpha$,
however, the situation is different because change in hopping leads
to more significant changes in the spectrum, as compared to the change
in $U$. Therefore, the dependence of the optical absorption on the
Hamiltonian parameters is itself dependent on the nature of the multilayer
ribbon concerned. Because, at present no experimental information
is available as to the optical absorption spectra of GNRs, thus, it
is impossible to judge as to what are the correct set of parameters
to be used in the Hamiltonian. Therefore, given our past experience
with the modified screened parameters ($U=6.0$ eV and $\kappa_{i,j}=2.0$
($i\neq j)$ and $\kappa_{i,i}=1$), and the in-plane hopping $t=2.7$
eV for the mono-layer GNRs,\cite{gundra} in the remainder of this
work we utilize modified screened parameters and the hopping values
$t=2.7$ eV ,$t^{'}=0.27$ eV, $t_{\bot}=0.4$ eV, $t_{\bot}^{'}=0.3$
eV.}

\subsection{Effect of edge alignment on the absorption spectrum}

\begin{figure}
\includegraphics[width=6cm]{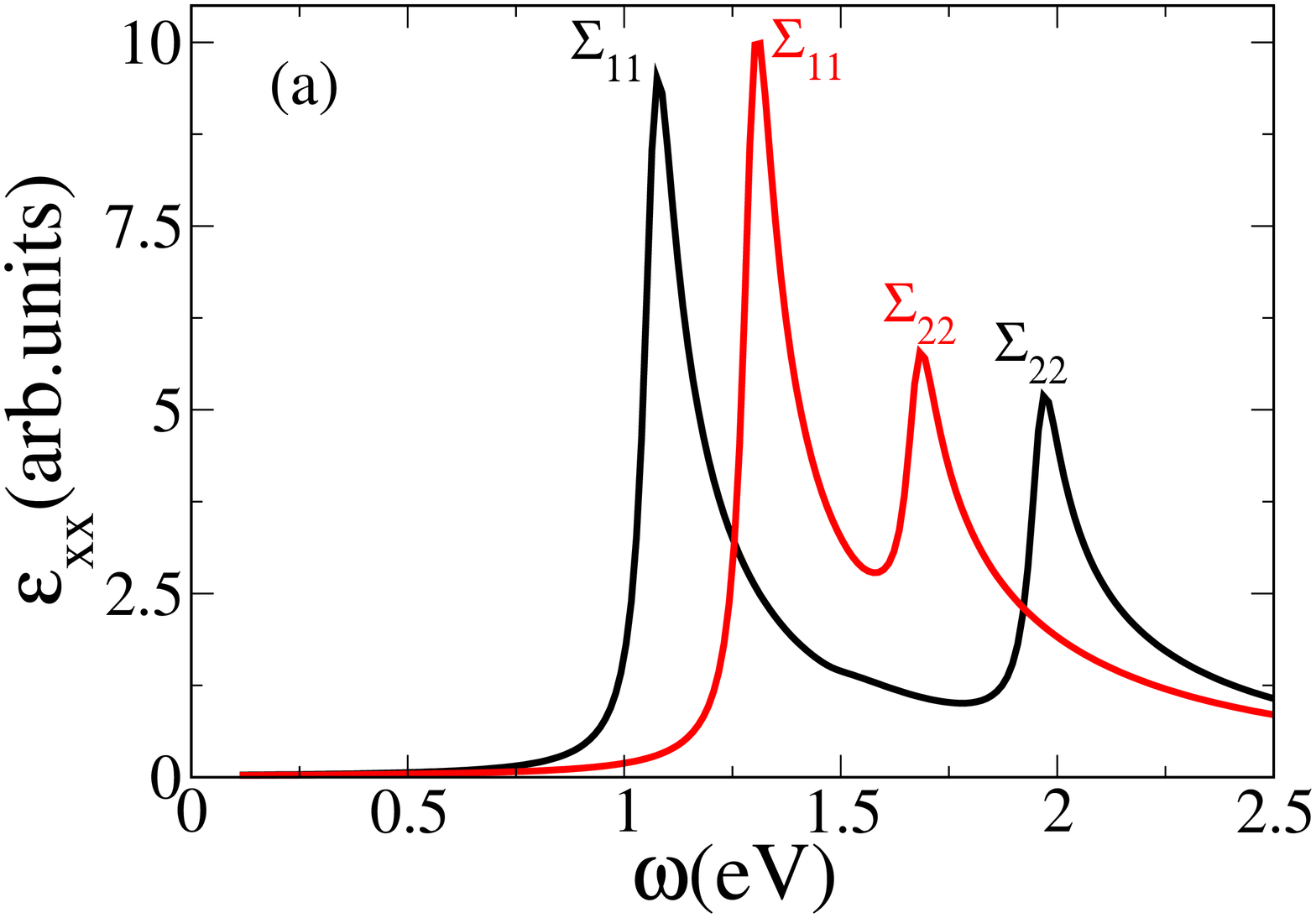}
\vspace*{3mm}
\includegraphics[width=6cm]{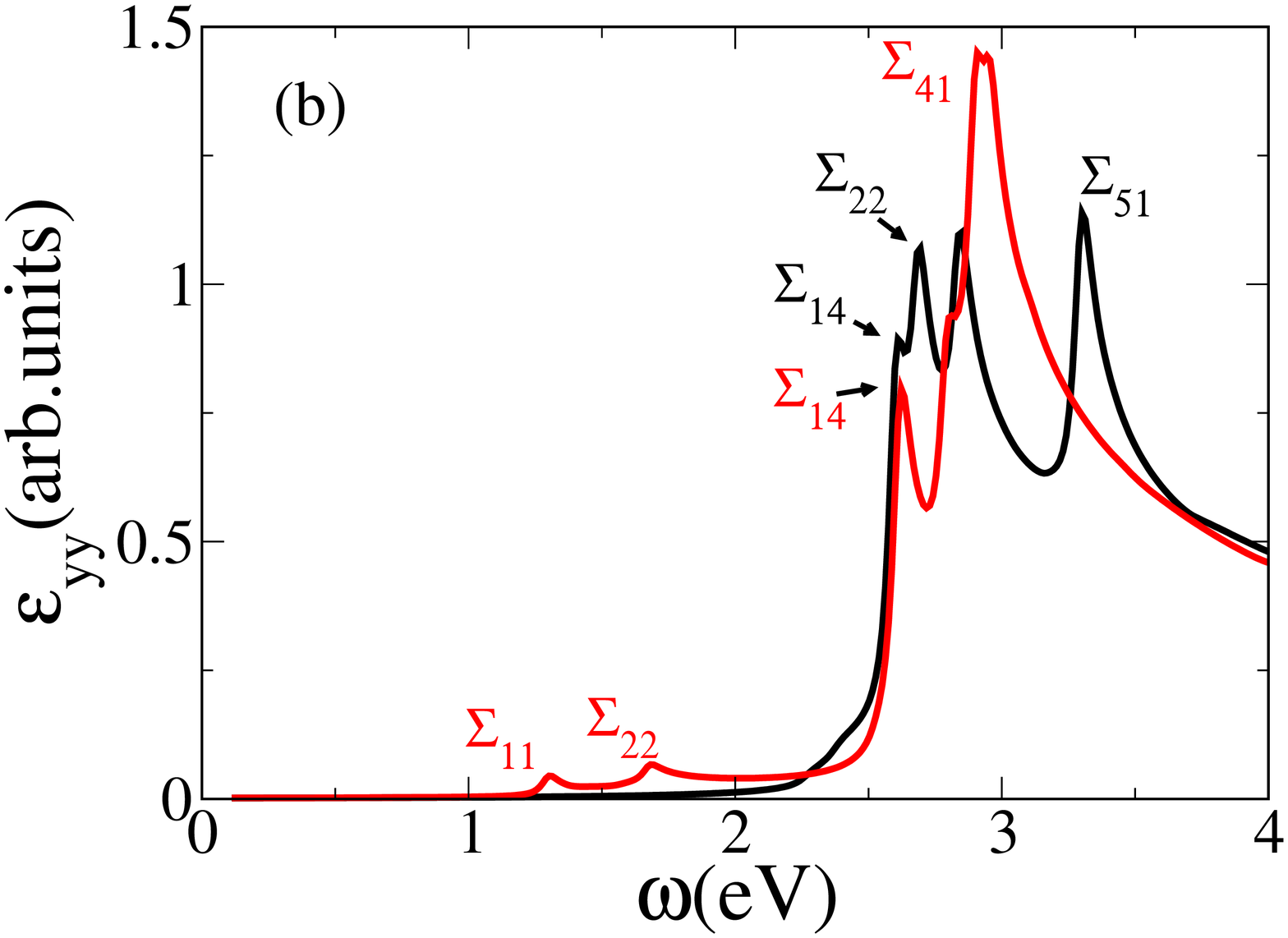}
\vspace*{3mm}
\includegraphics[width=6cm]{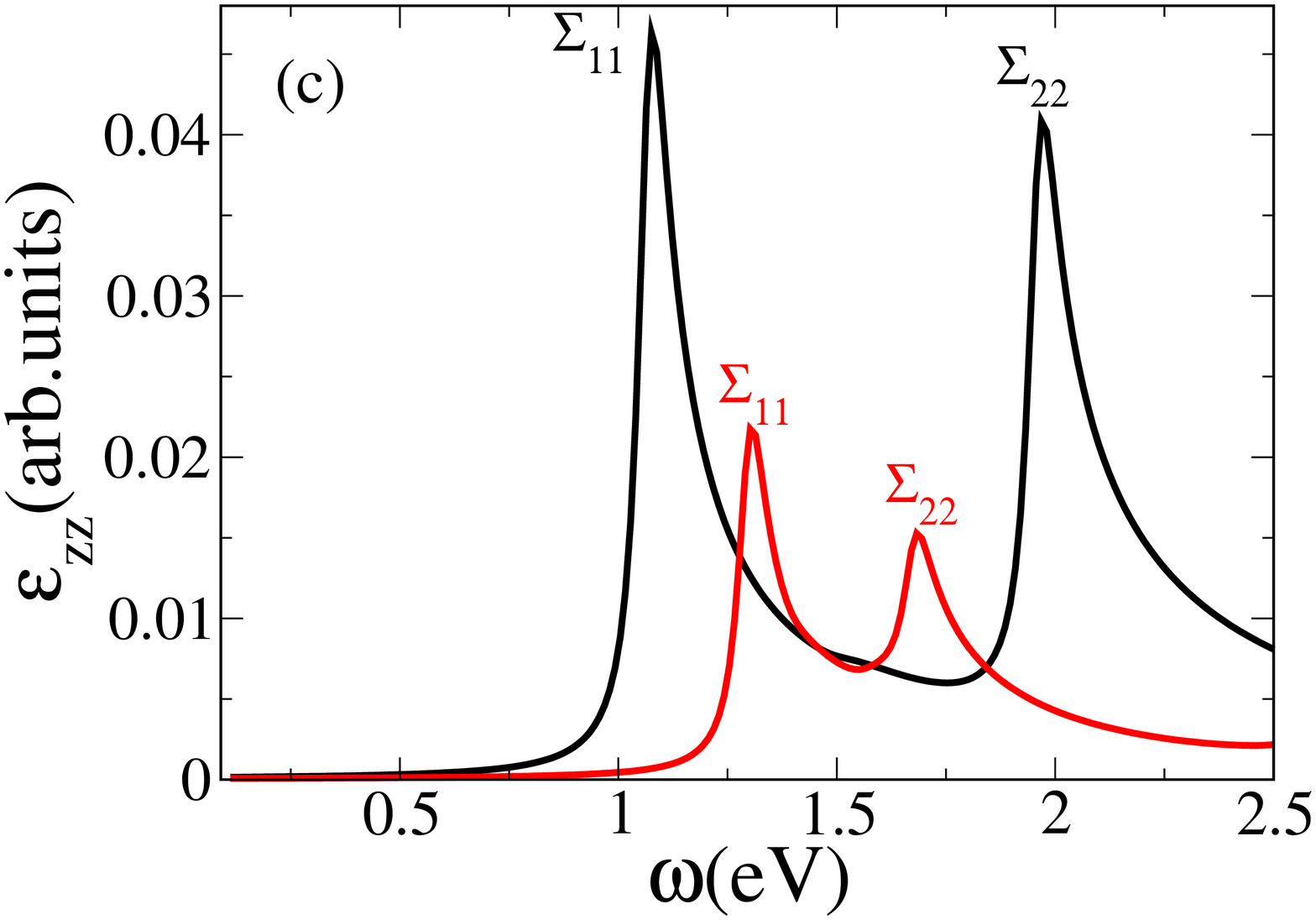}

\caption{(Color online) Optical absorption spectra of 2-AGNR-8 calculated using
the PPP-RHF approach for the light polarized along: (a) $x$ axis
(b) $y$ axis and (c) $z$ axis. The black lines represent the $\alpha$
alignment case, while the red lines denote the $\beta$ alignment.
A line width of 0.05 eV was assumed throughout.}
 \label{Flo:sig-alpha-beta-a8}
\end{figure}

In order to investigate the sensitivity of optical absorption to the
type of edge alignment, we present the optical absorption spectra
of 2-AGNR-8 in $\alpha$ and $\beta$ alignments. In Fig. \ref{Flo:sig-alpha-beta-a8}
a we present the optical absorption spectrum for light polarized along
the periodicity direction ($x$ axis, $\epsilon_{xx}(\omega)$), where
$\Sigma_{mn}$ denotes a peak in the spectrum due to a transition
from $m$-th valence band (counted from the top) to the $n$-th conduction
band (counted from the bottom). For $\alpha$-alignment (black line),
the first peak of $\epsilon_{xx}(\omega)$ at 1.06 eV is $\Sigma_{11}$,
while the one at 1.96 eV represents $\Sigma_{22}$. For the $\beta$
alignment (red line) $\Sigma_{11}$ peak is located at 1.30 eV, while
the second peak ($\Sigma_{22}$) at 1.67 eV. The separation between
the first two peaks is larger in the case of $\alpha$-alignment,
as compared to that in the $\beta$ alignment. This is a consequence
of the observation discussed in Sec. \ref{sub:bands} that the energy
separation between the adjacent bands near Fermi energy at $k=0$
is larger in the $\alpha$-alignment compared to $\beta$ -alignment. 

In Fig. \ref{Flo:sig-alpha-beta-a8}b, the absorption spectra corresponding
to the $y$-polarized photons ($\epsilon_{yy}(\omega)$) is presented,
and the first peak for the $\alpha$ alignment (black line) located
at 2.61 eV represents $\Sigma_{14}$, the second peak at 2.69 eV is
$\Sigma_{22}$, while the peak at 3.3 eV is $\Sigma_{51}$. Significant
differences in $\epsilon_{yy}(\omega)$ are observed for the $\beta$
alignment: it starts with two tiny peaks, with the first one ($\Sigma_{11}$)
at 1.30 eV, and the second one at 1.67 eV ($\Sigma_{22}$), which
were absent in case of the $\alpha$ alignment. The intense peaks
for $\beta$ -alignment are located at 2.62 eV ($\Sigma^{14}$), which
is coincidental with the same peak for the $\alpha$ case, and 3 eV
($\Sigma_{41}$+\textcolor{black}{$\Sigma_{23}$}) which is located
in the similar energy range as the higher peaks of the $\alpha$ case. 

Finally, we examine the $z$ polarized component of the absorption
spectra ($\epsilon_{zz}(\omega$)) presented in Fig. \ref{Flo:sig-alpha-beta-a8}c,
for 2-AGNR-8. In the low energy regime $\epsilon_{zz}(\omega)$ follows
$\epsilon_{xx}(\omega)$ very closely , but the magnitude of the peaks
of $\epsilon_{zz}(\omega)$ is reduced significantly compared to the
peaks of $\epsilon_{xx}(\omega)$, \textcolor{black}{because of the
weak inter-layer coupling.} For $\alpha$-alignment, the peak at 1.06
eV is $\Sigma^{11}$ and second peak at 1.96 eV is $\Sigma^{22}$,
while for the $\beta$ alignment, the corresponding peaks are at 1.30
eV, and 1.67, respectively. One very important difference in the spectra
of the two types of alignments is that the peak intensities in the
lower energy region are much higher in the $\alpha$-alignment, than
in the $\beta$-alignment. 

Therefore, the absorption spectra for each polarization direction
bring out distinct features depending on the type of edge alignment,
which can be used in the experiments on oriented samples to determine
the type of alignment.

\subsection{Variation of optical absorption with the number of layers }

\label{sub:opt-layers}

\begin{figure}
\includegraphics[width=6cm]{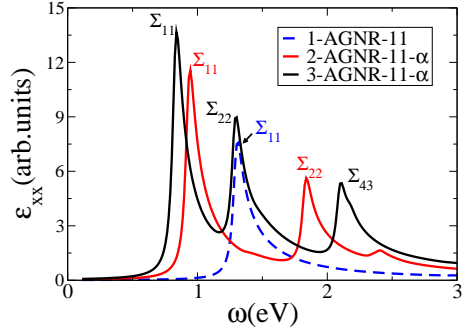}

\caption{(Color online) Optical absorption spectra of 1-AGNR-11-$\alpha$,
2-AGNR-11-$\alpha$ and 3-AGNR-11-$\alpha$ calculated using the PPP-RHF
approach, for the light polarized along the $x$ axis. A line width
of 0.05 eV was assumed throughout.}
\label{Fig:sig-vs-nl}
\end{figure}

Recently, Wright \emph{et al}.\cite{mGNR4-optics} using the TB calculations
performed on $3p+2$ family of bilayer AGNRs, showed that the inter-layer
coupling causes strong inter subband transitions, leading to tremendous
enhancement in the optical conductivity in the low-energy region,
thus, making them suitable candidates for opto-electronic applications.
In Fig. \ref{Fig:sig-vs-nl} we present the optical absorption spectra
for 1-AGNR-11, 2-AGNR-11-$\alpha$, and 3-AGNR-11-$\alpha$, calculated
using the PPP-RHF approach for the\textcolor{green}{{} }\textcolor{black}{$x$-polarized
light,} with the aim of understanding the influence of $n$ on the
optical absorption. In 1-AGNR-11, the $\Sigma^{11}$ peak of $\epsilon_{xx}(\omega)$
is located at 1.31 eV, while in 2-AGNR-11, and 3-AGNR-11 it becomes
more intense, and successively gets red shifted to 0.94 eV and 0.84
eV, respectively. A similar trend is observed also for other transition
peaks as well, as well as for the peaks in $\epsilon_{yy}(\omega)$
and $\epsilon_{zz}(\omega)$ (\textcolor{black}{figures not shown}).
Although, the band gaps obtained using the PPP-RHF approach are much
larger than that using the TB method, yet the tendency towards the
enhancement of the optical response with the increasing value of $n$
in the lower energy region is obvious from these plots. This behavior
is fully consistent with the earlier discussion in Sec. \ref{sub:gaps}
that the enhancement of the optical response in the lower energy region
of multilayer AGNRs can be understood in terms of the redshift of
the fundamental gap, and the increase in the joint density of states
near $E_{F}$, with increasing $n$. 

\subsection{Effect of gate bias on the optical absorption}

\begin{figure}
\vspace*{3mm}
\includegraphics[width=6cm]{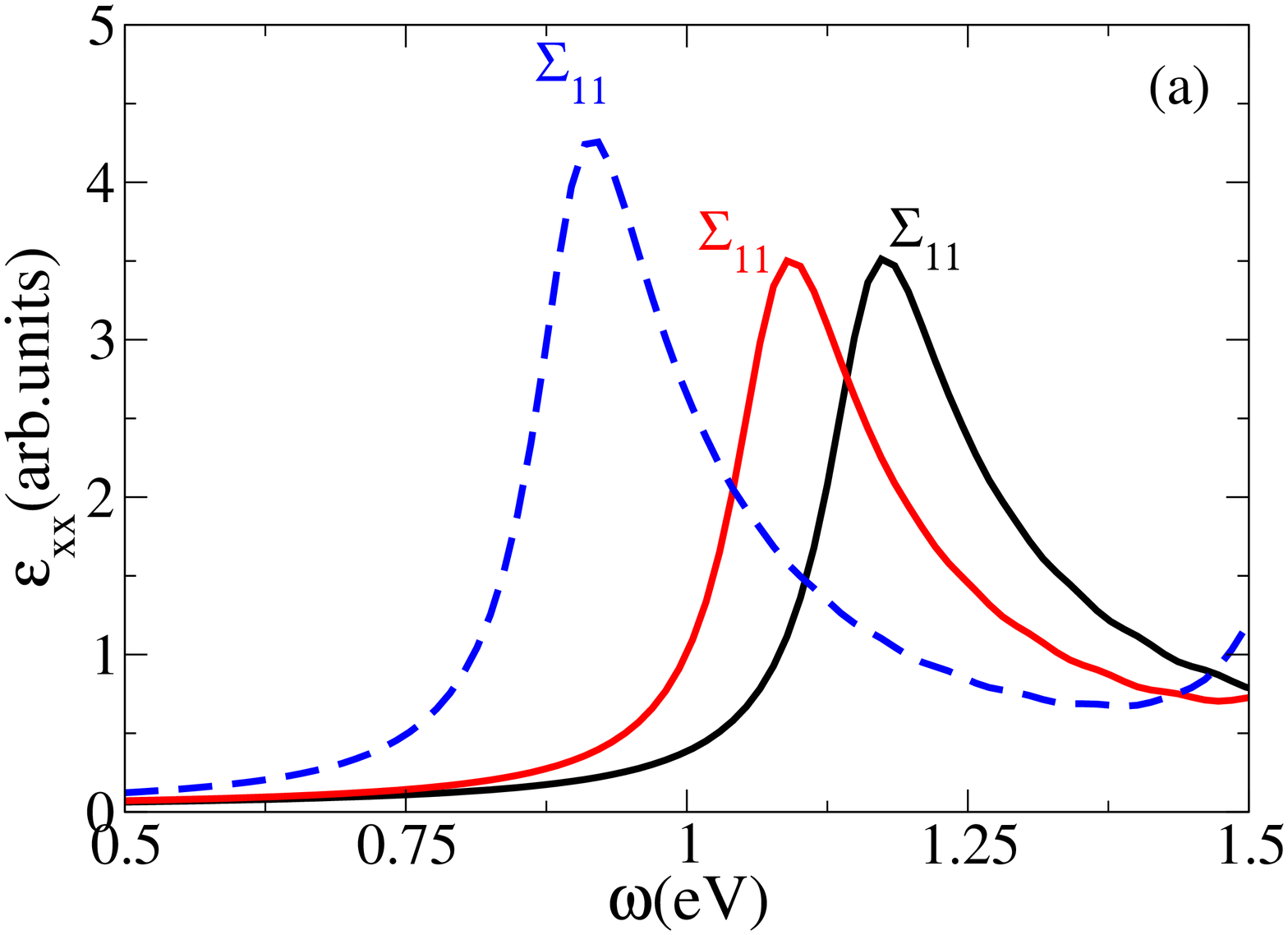}
\vspace*{3mm}
\includegraphics[width=6cm]{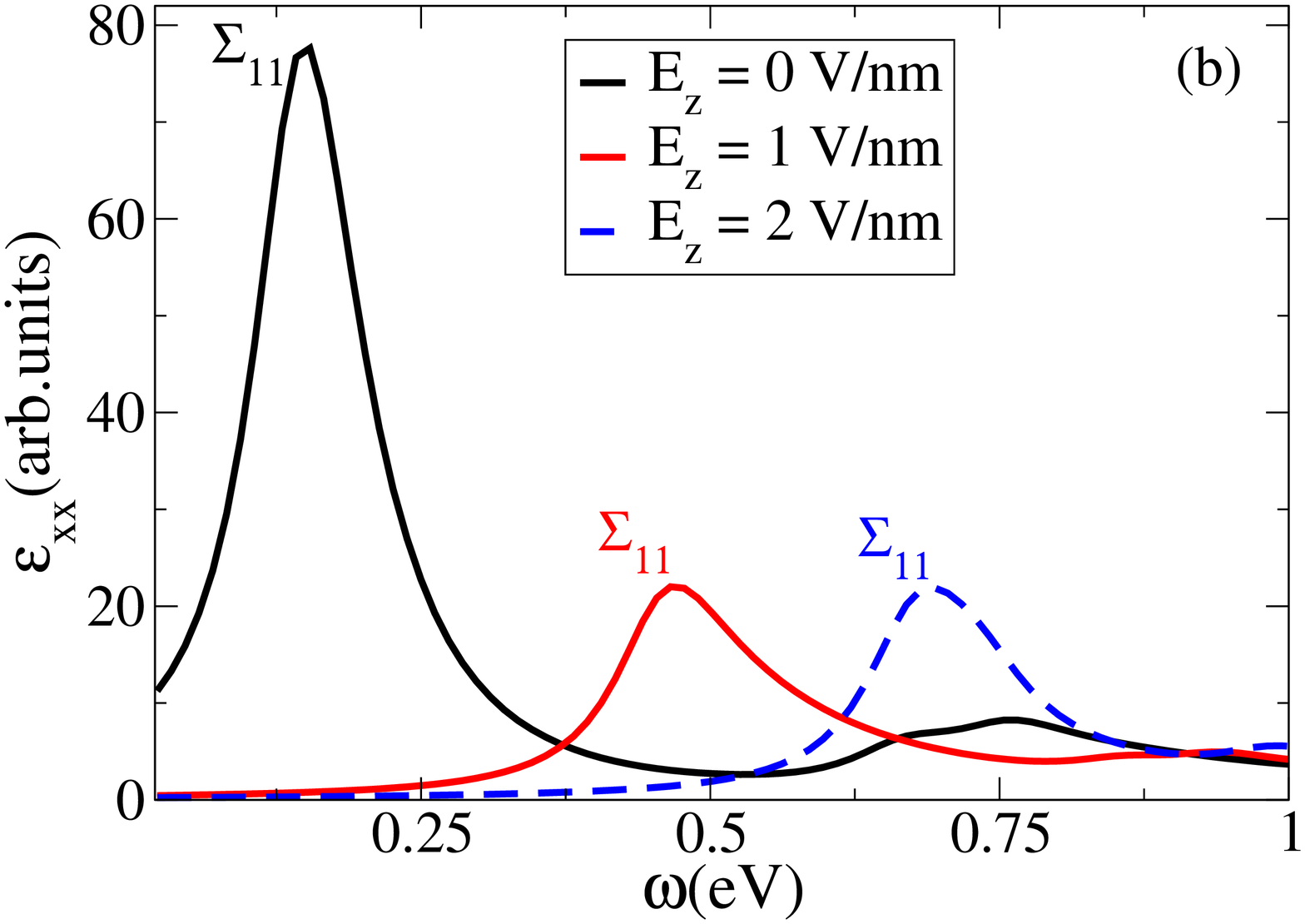}

\caption{(Color online)PPP-RHF absorption spectra at several values of the
static electric field $E_{z}$, for the incident light polarized along
the $x$ axis for (a) 2-AGNR-12-$\alpha$, (b) 2-AGNR-24-$\alpha$.
A line width of 0.05 eV assumed throughout.}

\label{Flo:sigx-vs-ez}
\end{figure}

\begin{figure}
\includegraphics[width=8cm]{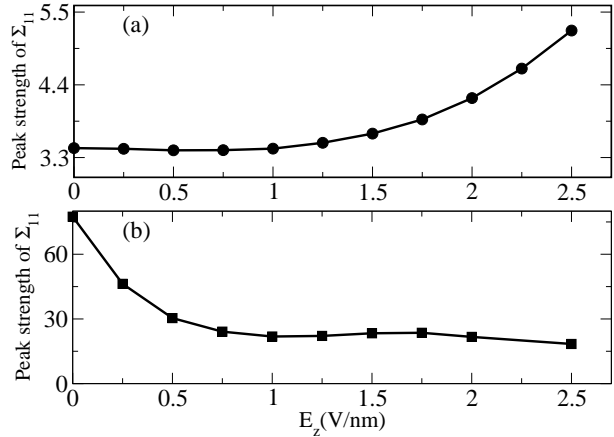}

\caption{Variation of peak strength of $\Sigma_{11}$ transition of $\epsilon_{xx}$
with static electric field $E_{z}$, computed using the PPP-RHF method,
for: (a) 2-AGNR-12-$\alpha$ (b) 2-AGNR-24-$\alpha$.}

\label{Flo:peak-strength}
\end{figure}

As discussed in Sec. \ref{sub:bands}, a gate bias has profound effects
on the electronic structure of multilayer AGNRs, and, therefore, it
is of great interest to understand the effect of electric field on
the optical properties of these systems. In the Fig.\ref{Flo:sigx-vs-ez}
a and \ref{Flo:sigx-vs-ez} b we present the absorption spectrum for
the $x$-polarized light ($\epsilon_{xx}(\omega)$) for 2-AGNR-12-$\alpha$
\textcolor{black}{and the 2-AGNR-24-$\alpha$ }respectively, for different
values of $E_{z}$. Upon increasing the value of $E_{z}$, we observe
the following trends in the plot. In 2-AGNR-12-$\alpha$, peak $\Sigma_{11}$,
which corresponds to the fundamental gap, gets red shifted, without
much change in the intensity at $E_{z}=1$V/nm. But, for $E_{z}=2$
V/nm, besides exhibiting the redshift, the peak also becomes more
intense. In case of 2-AGNR-24-$\alpha$, whose gap is less than $\epsilon_{c}$,
the $\Sigma_{11}$ peak expectedly blue shifts with increasing $E_{z}$,
however, it loses significant intensity as the field is increased
from $E_{z}=0$ to $E_{z}=1$ V/nm. However, upon further increasing
the field to $E_{z}=2$ V/nm, no significant change in the intensity
is observed. Thus, as far as the intensity of $\Sigma_{11}$ peak
with increasing gate bias is concerned, 2-AGNR-12-$\alpha$ \textcolor{black}{and
2-AGNR-24-$\alpha$ behave quite differently. To elucidate this point
further, in Fig.\ref{Flo:peak-strength} we present the variation
of the heights of the the} $\Sigma_{11}$ peaks of 2-AGNR-12-$\alpha$
and 2-AGNR-24-$\alpha$, as functions of the gate field, in the range
$0\leq E_{z}\leq2.5$ V/nm. Recalling that the height of a peak is
proportional to the oscillator strength of the corresponding transition,
we observe the following trends: (a) for 2-AGNR-12-$\alpha$ the intensity
varies slowly for $0\leq E_{z}\leq1.0$ V/nm, thereafter it increases
monotonically, while (b) for 2-AGNR-24-$\alpha$ the intensity decreases
rapidly for $0\leq E_{z}\leq1.0$ V/nm, and, thereafter it exhibits
slow oscillations around a constant value. Thus, the two ribbons exhibit
opposite behavior as far as the dependence of the oscillator strength
of $\Sigma_{11}$ peak on the bias field is concerned. This behavior
is in sharp contrast to the experimental\cite{Zhang-2,Castro} and
theoretical\cite{gated-bilayer-theory-yang} results on bilayer graphene,
which report a monotonic increase in the intensity of the first peak,
with an increasing gate bias. This peculiar behavior of the peak intensity
as a function of the gate bias can be tested in oriented samples of
multilayer AGNRs, and can be used to distinguish between the widths
of such ribbons.

\section{Summary and Outlook}

\label{sec:summary}

In summary, we have used the PPP model based $\pi$-electron approach,
incorporating long-range Coulomb interactions, to study the electronic
structure and optical properties of multilayer AGNRs, at the mean-field
Hartree Fock level. We considered two types of edge alignments namely
$\alpha$ and $\beta$, both in Bernal (AB) stacking pattern, and
systematically studied the variation of band gaps and optical absorption
spectra of multilayer AGNRs with the increasing number of layers.
These calculations have demonstrated anisotropic optical response
of multilayer AGNRs, with the calculated absorption spectra being
crucially dependent on the polarization direction of the incident
light. Furthermore, the optical response has been shown to depend
upon the type of the edge alignment ($\alpha$ vs. $\beta$) and the
number of layers, leading to the \textcolor{black}{possible} experimental
determination of the nature of the edge alignment, and the number
of layers in the system, using optical probes. Our calculations also
reveal that a gate bias along the inter-layer direction not only alters
the peak positions in the absorption spectra due to the change in
the band gaps, but also the peak intensities (oscillator strengths)
in a nontrivial way. The variation of the peak intensities as a function
of the gate bias is of the opposite nature for the narrow and the
broad bilayer AGNRs, a behavior in sharp contrast to bilayer graphene.
These predictions of ours can be tested in optical experiments performed
on the oriented samples of multilayer AGNRs.

As far as the future studies on multilayer AGNRs are concerned, it
will be interesting to probe the influence of electron-correlation
effects on various properties studied here. Furthermore, the nature
of excitons in multilayer AGNRs should be investigated, so as to provide
a complete description of the linear optical absorption spectra of
multilayer AGNRs. Calculations along those directions are underway
in our group, and results will be reported in future publications.
\begin{acknowledgments}
We thank the Department of Science and Technology (DST), Government
of India, for providing financial support for this work under Grant
No. SR/S2/CMP-13/2006. K. G. is grateful to Dr. S. V. G. Menon (BARC)
for his continued support of this work.\end{acknowledgments}

\end{document}